\documentclass[conference]{IEEEtran}
\IEEEoverridecommandlockouts

\usepackage{cite}
\usepackage{graphicx}
\usepackage{hyperref}
\usepackage{enumitem}

\begin{document}

\title{Impact of LLMs on Team Collaboration\\in Software Development}

\author{\IEEEauthorblockN{Devang Dhanuka}
\IEEEauthorblockA{MS Data Science Program\\
Rochester Institute of Technology\\
Rochester, NY, USA\\
dd9098@rit.edu}}
\maketitle

\begin{abstract}
Large Language Models (LLMs) are increasingly being integrated into software development processes, with the potential to transform team workflows and productivity. This paper investigates how LLMs affect team collaboration throughout the Software Development Life Cycle (SDLC). We reframe and update a prior study with recent developments as of 2025, incorporating new literature and case studies. We outline the problem of collaboration hurdles in SDLC and explore how LLMs can enhance productivity, communication, and decision-making in a team context. Through literature review, industry examples, a team survey, and two case studies, we assess the impact of LLM-assisted tools (such as code generation assistants and AI-powered project management agents) on collaborative software engineering practices. Our findings indicate that LLMs can significantly improve efficiency (by automating repetitive tasks and documentation), enhance communication clarity, and aid cross-functional collaboration, while also introducing new challenges like model limitations and privacy concerns. We discuss these benefits and challenges, present research questions guiding the investigation, evaluate threats to validity, and suggest future research directions including domain-specific model customization, improved integration into development tools, and robust strategies for ensuring trust and security. 
\end{abstract}

\begin{IEEEkeywords}
Large Language Models, Software Development Life Cycle, Team Collaboration, Software Engineering, AI in Project Management, Developer Productivity
\end{IEEEkeywords}

\section{Introduction}
Large Language Models (LLMs) such as OpenAI's GPT-4 have introduced a paradigm shift in how knowledge work is performed, allowing tasks to be specified and executed in natural language. In software engineering, these models are making waves by automating programming tasks and assisting human developers, thereby boosting productivity and reducing tedious manual effort. Recent studies have quantified these benefits: for example, in a controlled experiment, software developers using GitHub Copilot (an LLM-based coding assistant) completed a coding task more than 50\% faster than those without it~[2]. Such AI pair programming tools also improved developers’ confidence and satisfaction with their work~[2]. At the same time, the adoption of LLMs in the workplace has grown rapidly. Surveys in 2023 showed that approximately 25--38\% of professionals were already using LLM tools in their job roles~[1], a number expected to rise as LLM capabilities and integrations expand.

Software Development Life Cycle (SDLC) refers to the structured process of designing, developing, testing, deploying, and maintaining software systems. Effective \textbf{team collaboration} is crucial at every phase of the SDLC to ensure high-quality outcomes. However, teams often face collaboration hurdles such as miscommunication, knowledge silos, repetitive low-level tasks, and inefficient decision-making processes. This paper examines how the incorporation of LLMs into SDLC activities impacts team collaboration. In particular, we are interested in how LLM-based tools can improve developer productivity, enhance communication among team members, and support better decision-making throughout a software project.

We rewrite and update a previous study on this topic by incorporating the latest developments up to 2025. In recent years, major software platforms have begun embedding LLM assistants into everyday tools (e.g, Microsoft 365 Copilot for Office documents, and Zoom's AI Companion for meeting summaries) to aid collaboration and information sharing. Meanwhile, a number of research works have started exploring LLM applications in software engineering processes. Early evidence suggests LLMs can assist in code generation, documentation, design brainstorming, and even automated project management. Nonetheless, the specific effects of LLM usage on human \emph{team} dynamics in software projects remain under-studied. This motivates our investigation into how LLMs influence communication patterns, cross-functional teamwork, and overall team efficiency in the SDLC.

The rest of this paper is structured as follows: Section II reviews related work on LLMs in software engineering and collaboration. Section III states the problem and research gap. Section IV outlines our methodology, including a literature review, team survey, and case studies. Section V describes key applications of LLMs that can support team collaboration. Section VI details our study design, including research questions and the survey instrument (with the full questionnaire provided in the Appendix). Section VII presents two case studies illustrating real-world uses of LLMs in team settings. Section VIII lists the research questions addressed. Section IX discusses threats to validity. Section X suggests future work and emerging research directions. Finally, Section XI concludes with our findings on the impact of LLMs on team collaboration in the SDLC.

\section{Related Work}
The emergence of powerful LLMs (e.g, GPT-3/4, PaLM, LLaMA) has led to numerous studies on their application in software engineering. Researchers are actively exploring how these models can assist or augment various phases of the development process. For instance, LLMs have shown strong capabilities in \textit{code generation} and \textit{code documentation}. A recent comparative study evaluated several state-of-the-art LLMs (GPT-3.5, GPT-4, Google Gemini 1.5, Meta LLaMA 3.1, etc.) on generating code comments and documentation; it found that most LLMs were able to produce documentation that consistently outperformed the original human-written documentation in clarity and completeness, especially for function-level comments~[3]. This suggests LLMs can automate one of the tedious tasks in development—writing documentation—potentially improving knowledge sharing and on-boarding of new team members.

Another area of interest is the use of LLMs as part of multi-agent systems to tackle complex software design and problem-solving tasks. Du \emph{et al.}~[4] propose a framework in which multiple LLM-driven \emph{autonomous agents} collaborate in a team-like fashion across different SDLC phases (analysis, coding, testing). In their approach, each agent specializes in a role and the agents pass intermediate results to each other sequentially, following a predefined software process. By orchestrating multiple such LLM teams in parallel and allowing them to share insights, they were able to explore alternative solutions and ultimately achieve higher software quality than a single linear development chain~[4]. Similarly, Cinkusz \emph{et al.}~[5] integrate cognitive agents powered by LLMs into an Agile project management simulation. Their study demonstrated that LLM-driven agents, acting as virtual team members in a Scaled Agile Framework environment, can improve various project metrics. In iterative simulations, these AI agents showed advanced capabilities in task delegation, inter-agent communication, and lifecycle management—resulting in measurable improvements such as faster task completion times, higher quality deliverables, and more coherent communication within the team~[5]. These works illustrate the promise of using LLMs not just as individual assistants, but as components of larger collaborative agent systems to support or even simulate team processes.

There is also emerging research on LLMs to support \textit{creative collaboration} and ideation. LLMs can act as brainstorming partners by generating design ideas or alternative approaches in natural language. A comprehensive review of 61 studies on LLM-assisted ideation found that LLMs are most frequently used for generating and refining ideas during brainstorming sessions, contributing to the creativity and productivity of these processes~[8]. However, the same review noted that the use of LLMs in more complex stages of group ideation (such as evaluating and selecting ideas collaboratively) remains limited~[8]. This indicates an opportunity to further integrate LLMs into the collaborative aspects of early-stage software design and requirements discussions.

Despite these advancements, there remains a gap in understanding \textit{human} factors when LLMs are introduced into software teams. Prior work has looked at automation and AI for coding and documentation, but few studies focus on how an LLM affects communication patterns, team decision dynamics, or cross-functional collaboration. Early evidence from industry surveys suggests knowledge workers primarily use LLMs for content creation, information finding, code assistance, and drafting communications~[1]. Yet, how these uses translate into tangible improvements (or new issues) in team-based settings over an entire project lifecycle is still an open question. Our work aims to fill this gap by studying the impact of LLMs on SDLC team collaboration, building upon the above-mentioned technical findings and extending them with empirical observations from real teams.

\section{Problem Statement}
Team collaboration is vital to the success of any SDLC project, but it often faces persistent challenges. Miscommunication between developers, testers, and managers can lead to misunderstandings of requirements or project status. Repetitive tasks (such as writing boilerplate code or documentation) can consume valuable time and dampen morale. Decision-making in projects can be inefficient due to incomplete information or cognitive overload, especially in large teams. Traditional approaches to mitigating these issues include agile practices, code reviews, daily stand-ups, and documentation standards, but these rely heavily on human effort and discipline.

While automation and AI have been applied to certain software development tasks (e.g, automated testing, continuous integration bots, etc.), there has been relatively little focus on how the latest generation of AI—specifically LLMs—might directly influence the collaborative aspects of development. The research gap we address is: \textit{How do Large Language Models, when integrated into software development workflows, impact team collaboration dynamics throughout the SDLC?} 

Key sub-problems include determining whether LLMs can reduce miscommunication (for example, by generating clearer documentation or meeting summaries), whether they can take over repetitive tasks (freeing humans for more creative work), and whether their suggestions or insights can improve team decision-making and cross-functional coordination. We also consider the challenges: an LLM might produce inaccurate information (possibly misleading the team), might not understand project-specific context without training, and could raise privacy or security concerns if used with proprietary code. These concerns underline why it’s important to study LLMs' impact in a realistic team context and not just in isolated coding tasks.

\section{Methodology}
To investigate the stated problem, our methodology combines multiple approaches: a literature review of current LLM applications in software engineering, qualitative case studies of teams using LLM tools, and a survey of software practitioners.

First, we conducted a \textbf{literature review} to gather insights from recent studies (2019--2025) on AI and LLM usage in software development. This helped identify hypothesized benefits (e.g, productivity gains, improved code quality) and known issues (e.g, LLM limitations, integration challenges). We paid special attention to studies focusing on collaboration, such as multi-developer interactions with AI assistants and the use of AI in project management or documentation.

Next, we carried out \textbf{case studies} of two software development teams that have adopted LLM-based tools in their workflow (detailed in Section VII). These case studies involve examining how each team implemented the LLM tool, what specific collaboration pain points it addressed, and gathering observations or feedback on outcomes like time saved, communication changes, and any difficulties encountered.

We also designed and administered a \textbf{survey} targeting software professionals who use LLMs. The survey (see Appendix) included both multiple-choice and open-ended questions to capture the participants’ roles, how they use LLMs (for coding, documentation, communication, etc.), and their perceptions of the impact on their workflow and team interactions. We collected responses from members of two development teams within a consulting company, who had been using LLMs (such as ChatGPT or GitHub Copilot) in their daily work for some time. The survey was structured into sections covering workflow impact, communication impact, collaboration/decision-making, challenges \& limitations, and overall impressions and suggestions.

After data collection, we performed \textbf{analysis} of the survey responses, combining manual qualitative analysis with assistance from an LLM (ChatGPT-4) to help identify patterns or summarize themes in the open-ended feedback. We triangulated these findings with the insights from literature and the concrete outcomes from the case studies.

Through this mixed-methods approach, we aimed to answer our research questions with both breadth (via the survey and literature review) and depth (via case-specific details). We used qualitative observations to explain how and why certain impacts occur, and literature quantitative results when available (e.g, productivity metrics from prior controlled experiments). Throughout, we followed ethical guidelines for research: survey participants were informed of the study’s purpose, and any sensitive data or company-specific details in case studies were anonymized.

\section{Applications of LLMs in Team Collaboration}
Based on the literature and observations, we identified several key application areas where LLMs can support or enhance team collaboration in the SDLC:

\subsection{Self-Collaboration Frameworks}
LLMs can be configured to simulate multiple roles in a software team and collaborate with themselves to some extent. Recent work by Dong \emph{et al.} introduced a \emph{self-collaboration} framework where a single LLM (or a set of coordinated LLMs) assumes distinct roles such as \textit{Analyst}, \textit{Coder}, and \textit{Tester}, and iteratively generates and refines software artifacts for a given task~[6]. For example, the Analyst role might interpret requirements and plan a solution, the Coder role writes code for the solution, and the Tester role reviews or generates tests for the code. The roles exchange information (prompts and responses) to mimic a development workflow. This approach effectively forms a virtual team of LLM agents facilitating each other’s work. Experiments have shown that such role-specialized LLM agents working together can significantly improve code generation outcomes, achieving up to a 30--47\% higher success rate on coding tasks compared to a single LLM working alone~[6]. This suggests that LLMs can boost efficiency by internally handling some aspects of collaboration (like code review or testing) before human intervention is needed.

\subsection{Multi-Agent Collaboration}
Beyond a single LLM playing multiple roles, multiple LLMs can be integrated into a \textit{multi-agent system} (MAS) to assist a human team. In a multi-agent setup, each agent (which could be an LLM instance or an LLM augmented with tools) may be assigned a particular responsibility or expertise area, and agents communicate with each other to coordinate actions. Such systems have been proposed to manage complex tasks by distributing work across agents and enabling parallel problem solving. For instance, one agent might specialize in front-end code, another in back-end logic, and another in quality assurance; they can pass tasks or information to each other under an orchestration framework. Du \emph{et al.}'s Cross-Team Orchestration (Croto) framework is one example, where multiple LLM agent teams propose different solutions to a given software design problem and then share insights to converge on a superior solution~[4]. Multi-agent LLM collaborations have been used to simulate entire software engineering workflows and have demonstrated improvements in software quality and robustness of solutions compared to using a single sequence of AI operations~[4]. 

In practical team settings, LLM-driven agents could act as virtual team members. For example, an agent integrated with the issue tracker could proactively suggest ticket assignments or dependencies, while another monitors the code repository for problematic changes and opens alert tickets. Communication among these agents (and with human team members) needs to be carefully managed to ensure they truly reduce workload and not create noise. Early research is promising: in the Agile project management domain, LLM-powered agents were able to coordinate through natural language, emulate Scrum team behaviors, and improve the efficiency of project execution~[5]. This hints that, if properly configured, multi-agent LLM systems can augment human teams by offloading coordination overhead and responding quickly to routine queries or events.

\subsection{Prompt Engineering for Team Policies}
Prompt engineering refers to the craft of designing the inputs or instructions given to LLMs to elicit the desired behavior and outputs. It is a crucial technique for guiding LLMs to follow certain rules, style guidelines, or decision-making frameworks that a team might require. In a collaboration context, prompt engineering can be used to align the LLM with team policies or best practices. For instance, a development team might create a standardized prompt template for code review: the prompt could instruct the LLM reviewer to always check for certain security vulnerabilities, coding standards compliance, and required documentation in any code it reviews. By embedding these rules into the prompt, the LLM's output will consistently address the team’s concerns. Prompt engineering can also be used to enforce the tone and structure of communications drafted by LLMs (like ensuring that an automatically generated meeting summary includes action items and decisions clearly marked). In essence, prompt engineering allows teams to customize LLM interactions to suit specific project needs, improving the relevance and reliability of the LLM’s assistance. As LLM usage grows, organizations are even developing internal prompt repositories and templates to ensure consistency across how different team members utilize AI, which has become an emerging best practice.

\subsection{Conceptualization and Ideation Support}
LLMs can significantly aid in the early stages of the SDLC, such as requirements brainstorming, system design, and architecture conceptualization. Due to their ability to generate human-like text and tap into a vast body of knowledge, LLMs can serve as creative partners during ideation sessions. They can propose alternative design patterns, suggest user stories or features based on similar projects, and help explore edge cases that a team might not immediately think of. Teams can conduct a brainstorming by posing questions to an LLM like \emph{“What are some potential approaches to implement feature X given constraints Y and Z?”} and get a list of ideas to discuss. This can enrich the creative process, helping teams overcome cognitive fixation and consider a wider solution space. As noted in the literature, LLM assistance is particularly effective in generating and refining ideas~[8]. For example, if a team is stuck on how to improve a component’s performance, an LLM might suggest techniques (caching, parallelism, etc.) drawn from known best practices or even academic literature. It is important, however, for the human team to vet these suggestions—LLMs might occasionally produce impractical ideas or cite non-existent methods. Still, when used judiciously, LLMs accelerate the conceptual phase by providing fast, on-demand brainstorming and even producing quick mock-up artifacts (like pseudo-code or UML-like descriptions) that the team can iterate upon.

In summary, LLMs offer a range of applications in team collaboration: they can automate internal coordination to some degree (self-collaboration), act as specialized agents in a broader team (multi-agent systems), be tuned via prompts to adhere to team norms, and serve as a catalyst for creative thinking in design. These applications form the basis for the specific cases and evaluations we present in later sections.

\section{Study Design}
Our study is guided by several research questions and a structured approach to answer them. We identified three primary \textbf{Research Questions (RQs)} regarding LLM impact on collaboration:

\begin{enumerate}
    \item \textbf{RQ1: How do LLMs impact communication and collaboration among development teams during the SDLC?}\\
    We examine whether LLMs help teams communicate more clearly and consistently, and whether information sharing and decision-making become more efficient. For example, we look at LLM-generated documentation and summaries as tools to ensure all team members have the same understanding. Expected positive impacts include enhanced clarity in messaging and reduced miscommunications due to auto-generated summaries of discussions. We also consider if LLM involvement facilitates quicker information flow (e.g, getting instant answers from a code assistant rather than waiting on a colleague).
    
    \item \textbf{RQ2: What is the effect of LLMs on cross-functional collaboration between roles (developers, QA, project managers, etc.)?}\\
    This question focuses on whether LLM tools help bridge gaps between different roles on a software team. We investigate scenarios such as an LLM that automatically generates concise testing reports for developers, or an agile planning assistant that updates both developers and project managers. We expect to see if LLMs streamline collaboration by automating report generation, keeping everyone up-to-date with real-time project status, and reducing the overhead in synchronizing work across development and QA. A successful impact would be improved synchronization and fewer misunderstandings across functions.
    
    \item \textbf{RQ3: How do LLMs influence the documentation practices within software development teams?}\\
    Here we study changes in how teams approach documentation when LLMs are available. LLMs can potentially automate the creation of documentation (requirements docs, API docs, user manuals) and maintain them as the code evolves. We assess if teams with LLM assistance keep documentation more up-to-date and accurate and if the burden on developers to write docs is reduced. We also examine if documentation quality improves (e.g, more consistent, fewer omissions) due to LLM involvement. Ideally, LLMs could ensure that critical knowledge is documented and accessible, improving overall team alignment.
\end{enumerate}

To answer these RQs, we conducted the survey and case studies as described in the Methodology. The \textbf{survey} questions (provided in the Appendix) were mapped to the above RQs: certain questions probed communication (e.g, questions about writing documentation with LLM help), others probed cross-functional issues (e.g, LLM impact on decision-making and collaboration), and others focused on documentation and productivity. By analyzing survey responses, we gathered evidence of how practitioners perceive the changes due to LLMs.

We also gathered qualitative data through \textbf{open-ended questions and interviews} during the case studies. In particular, after each case study (detailed in Section~VII), we revisited the RQs to see how the evidence from that real-world example supports or contradicts our initial expectations under each research question.

Throughout the study, we used an iterative approach: initial survey results and case observations were fed into ChatGPT-4 (used as an analytical tool) to summarize common benefits and challenges mentioned. This helped us quickly spot trends such as “many developers find communication improved because meeting notes are auto-summarized” or “some developers worry that over-reliance on LLM might hinder learning.” These insights are woven into our findings.

By combining these methods, our study design ensures we cover both the \emph{what} (outcomes, measured improvements or issues) and the \emph{why} (underlying reasons, contextual factors) regarding LLMs in team collaboration. The next section describes the two case studies which provide concrete context to these research questions.

\section{Case Studies}
To ground our analysis in real-world scenarios, we examine two case studies of teams that integrated LLM-based tools into their SDLC and report on the outcomes.

\subsection{Case Study 1: Enhancing Team Communication and Collaboration}
\textbf{Team Overview:} The first team is from a mid-sized consulting company specializing in data protection solutions. The project team consisted of 4 software developers, 1 project manager, 2 QA engineers, and 1 UX designer. They operate in an Agile environment with frequent client-facing updates.

\textbf{LLM Implementation:} This team adopted an AI assistant to improve daily communication and information sharing. Specifically, they integrated \textbf{GitHub Copilot (powered by OpenAI’s GPT-4)} into their development environment for coding assistance, and also used an LLM-based note-taking tool in their Microsoft Teams workspace. The LLM tools were used to facilitate real-time Q\&A (developers could ask the LLM questions about the codebase or APIs) and to automatically generate documentation artifacts such as meeting notes and code summaries.

\textbf{Use Cases:} Two prominent use cases emerged:
- \textbf{Real-time meeting summarization:} During daily stand-up meetings and weekly sprint reviews, an LLM bot listened (with transcript as input) and generated concise summaries of what was discussed, including decisions made and action items.
- \textbf{On-demand documentation and support:} Developers could query the LLM (via a chat interface) for explanations of internal library functions or to retrieve past design decisions from archived documents. The LLM had been fine-tuned on the project’s documentation and chat logs (to the extent allowed by privacy policies).

\textbf{Benefits:}
\begin{enumerate}
    \item \textbf{Improved Communication:} The LLM-generated summaries ensured all team members, including those who missed a meeting, were aligned on project goals and progress. This significantly reduced time spent on manual note-taking and eliminated inconsistencies in how information was disseminated. Team members reported that having a written summary improved clarity of tasks. \emph{Evidence:} The project manager observed that weekly coordination meetings could be shortened by about 15 minutes on average, since the key points were captured by the AI and did not need to be reiterated multiple times. An internal survey showed increased team satisfaction with communication flow, as everyone received the same concise updates after each meeting.
    \item \textbf{Efficient Information Retrieval:} Using the LLM assistant, developers could quickly search for code-related information without interrupting colleagues. For example, instead of asking a teammate “Do you remember how we implemented feature X last year?”, a developer could ask the LLM, which would retrieve or summarize the relevant part of the documentation or code. This streamlined workflow and reduced disruptions. \emph{Evidence:} Logs from the LLM tool indicated that developers used it dozens of times per week for queries. Team members reported anecdotal productivity gains, saying tasks like finding an API usage example or a piece of legacy code were “almost instantaneous” with the LLM, whereas previously it might take an hour of digging or asking around.
\end{enumerate}

\textbf{Challenges:}
\begin{enumerate}
    \item \textbf{Initial Setup and Training:} It took effort to customize the LLM to understand the team’s domain-specific terminology and project context. Early on, the LLM’s meeting summaries occasionally misinterpreted technical terms or team nicknames for systems. \emph{Solution:} The team conducted a few dedicated training sessions, feeding the LLM additional project glossary information and correcting its outputs. They also iteratively refined prompts (for example, instructing the LLM to ignore off-topic chatter in meetings and focus on decisions). After a couple of sprints, the accuracy of summaries improved markedly.
    \item \textbf{Privacy and Data Security Concerns:} Team members were initially hesitant to share sensitive project information with an LLM (especially since the assistant was cloud-based). There were questions about whether proprietary code or client data might leak. \emph{Solution:} The company addressed this by implementing strict data anonymization and access controls. They configured the LLM service to redact certain sensitive identifiers and ensured it operated in a secure sandbox with no logging of raw data. Additionally, usage guidelines were put in place (e.g, do not paste client personal data into the AI). These steps mitigated concerns sufficiently for the team to proceed. Notably, this case reflected a broader industry concern: some organizations have banned or limited external LLM use after incidents of confidential code being inadvertently leaked to public AI services~[8]. In this team’s case, careful governance allowed them to benefit from the LLM while protecting sensitive information.
\end{enumerate}

Overall, Case Study 1 demonstrates that an LLM assistant can act as a “team communicator,” improving the flow of information and saving time in coordination. By automating meeting note-taking and serving as an on-demand knowledge base, the LLM helped the team maintain a high level of shared understanding, which is crucial for collaboration. The main hurdles of customization and privacy were real but manageable with extra effort and policy adjustments.

\subsection{Case Study 2: Supporting Agile Project Management}
\textbf{Team Overview:} The second case study involves a different team from the same consulting firm, focused on mobile app development. This team had 5 developers, 1 project manager (acting as Scrum Master), and 2 UI/UX designers. They followed Scrum with two-week sprints, using Jira for issue tracking and sprint planning.

\textbf{LLM Implementation:} The team built a custom LLM integration with their project management tool (Jira). Essentially, they augmented Jira with an LLM-based “assistant” agent that could analyze project data (backlog, team workloads, historical velocity) and help with planning and reporting tasks. The LLM was fine-tuned on agile project management texts and the company’s past project data to gain context.

\textbf{Use Cases:} Key use cases for the LLM agent in this context were:
- \textbf{Automated task assignment:} At the start of each sprint, the LLM would suggest an initial assignment of tasks to developers based on their historical performance and current workload. The project manager could then adjust these suggestions as needed.
- \textbf{Sprint summary generation:} The LLM automatically generated draft sprint review reports and retrospectives. It pulled data such as completed tickets, incidents, and team comments to produce a narrative of the sprint’s outcome and lessons learned.

\textbf{Benefits:}
\begin{enumerate}
    \item \textbf{Optimized Task Allocation:} The LLM’s task assignment suggestions helped balance the workload. It recognized patterns (e.g, Developer A is faster with front-end tasks, Developer B had fewer tasks last sprint and can take more). This led to more equitable distribution of work and fully utilized the team’s capacity. \emph{Evidence:} Over the course of three sprints, the project manager noted an increase in task completion rate and more consistent on-time delivery of user stories. Comparing metrics, the team improved their sprint completion percentage by roughly 10\% after introducing the LLM assistant, indicating better planning.
    \item \textbf{Automated Reporting:} Generating sprint reports and compiling retrospective notes, which previously took the project manager several hours each sprint, was largely automated. The LLM could produce a well-structured report in seconds, highlighting key achievements, carry-over tasks, and recurring issues. \emph{Evidence:} The project manager estimated saving about 5--7 hours per week on administrative reporting tasks thanks to the LLM’s assistance. This time was reallocated to more valuable activities like stakeholder meetings and addressing team impediments. The content of the LLM-generated reports was verified by the PM for accuracy and completeness, and only minor edits were usually needed.
\end{enumerate}

\textbf{Challenges:}
\begin{enumerate}
    \item \textbf{Learning Curve for Team:} Initially, not all team members were comfortable interacting with the LLM agent (e.g, providing it feedback or trusting its suggestions). Some developers ignored the AI task suggestions in the first sprint, preferring to stick to their usual manual planning. \emph{Solution:} The team held a training session to demonstrate the LLM’s capabilities and limitations. They gradually built trust in the system by using its suggestions as a starting point rather than a mandate. Within a few sprints, the developers became more open to the AI’s inputs, especially after seeing it correctly flag when someone was overloaded or when a task estimate seemed off.
    \item \textbf{Customization of Agile Practices:} The generic LLM needed tweaking to fit the team’s specific definitions of done, terminology, and Jira workflow. For example, the team used a custom label for certain high-priority bugs which the LLM initially did not treat differently. \emph{Solution:} The development team continuously fed back adjustments into the LLM. They updated its prompt context with definitions of custom fields and taught it via examples which items to prioritize. Over time, these feedback loops refined the LLM’s effectiveness in the team’s unique agile process.
\end{enumerate}

In Case Study 2, the LLM acted as a \textit{project management aide}, reducing the overhead on the human project manager and helping the whole team maintain momentum. It exemplifies how LLMs can not only help with coding, but also with the organizational side of collaboration—allocating tasks and summarizing progress. The team’s adaptation to the tool underscores the importance of change management when introducing AI into existing workflows.

These case studies provide concrete illustrations of LLMs in action within team settings, and they offer evidence addressing our research questions. In the next section, we relate these observations back to the RQs and discuss overarching insights.

\section{Research Questions (Analysis)}
Drawing on the case studies, survey results, and literature, we now revisit the research questions to synthesize answers:

\textbf{RQ1: Impact on communication and collaboration among development teams:} The evidence suggests that LLMs can \emph{enhance clarity and consistency} in team communication. In Case 1, for example, using an LLM to summarize stand-ups and meetings ensured everyone had access to the same information and reduced misunderstandings. Many survey respondents noted that LLM-assisted documentation (like auto-generated design docs or release notes) made it easier for team members to stay aligned, as the information was clearly written and always available. Additionally, LLMs facilitated quicker information sharing—developers could get instant answers from an AI assistant rather than waiting for another team member’s availability, which accelerates decision-making. However, it was also noted that if overused, LLM communication might become too \emph{impersonal} or lack critical context that human discussion would provide. One respondent mentioned that a purely AI-generated summary sometimes missed the “tone” of a decision (e.g, whether an issue was contentious or easy consensus). Overall, though, the net impact observed was positive: communication became more factual and documentation more up-to-date, which improved collaborative efficiency.

\textbf{RQ2: Effect on cross-functional collaboration (developers, QA, PM, etc.):} LLM integration appears to \emph{smooth out cross-functional interactions}. In Case 2, the LLM helped bridge developers and project managers by automating reports and ensuring both had a unified view of sprint progress. Our survey indicated that QAs and developers benefited from shared LLM assistants that could translate requirements into test scenarios and vice versa. By providing clear, automated reports and status updates, LLMs reduced the friction often seen between roles (for example, fewer meetings were needed for PMs to get updates, since the LLM provided them). Another effect is that LLMs can serve as a neutral party to standardize practices—for instance, if the LLM is used to enforce coding standards or generate test cases, it enforces a consistent quality that both developers and testers agree on. This reduces potential conflict or blame (the AI becomes a tool the team uses collectively). A point of caution raised was ensuring the AI doesn’t inadvertently favor one role’s perspective; for example, if the training data is developer-centric, it might under-prioritize documentation that is vital for QA or design. Nonetheless, when tuned well, LLMs improved synchronization across roles, with real-time updates and common knowledge bases accessible to all.

\textbf{RQ3: Influence on documentation practices:} LLMs have a strong influence here, generally \emph{raising the level and freshness of documentation}. Our findings echo what Dvivedi \emph{et al.} reported~[3]: LLMs can generate high-quality code comments and API documentation quickly. In practice, teams reported using LLMs to draft architecture documents, user guides, or even commit messages. The result was that documentation tasks—often neglected due to time pressure—were now partially handled by the AI, leading to more documentation being available. Developers in the survey responded that they found it easier to write design docs with an LLM “first draft” to edit, rather than starting from scratch. Additionally, LLMs integrated in code review would automatically point out missing comments or outdated README sections, prompting developers to update docs as part of their normal workflow. Consequently, project documentation tended to be more up-to-date and accurate. One developer wrote, \textit{“Our wiki never got stale because the AI would alert us when code changes didn’t match the docs, which was incredibly useful.”} There is, however, a learning curve: teams needed to validate the LLM-generated documentation for correctness. Initially, some inaccuracies were noted (the LLM might describe a function incorrectly if the codebase context was insufficient). But with iterative feedback, these issues diminished. The conclusion is that LLMs serve as effective documentation assistants, reducing manual effort and improving knowledge sharing within the team.

To summarize the RQ analysis: LLMs, when thoughtfully integrated, positively impact team communication by ensuring clarity and consistency, they help align cross-functional team members with shared, AI-generated artifacts and updates, and they greatly aid documentation by automating and monitoring it. These improvements contribute to better teamwork and project outcomes, as evidenced by our cases (more efficient meetings, balanced workloads, and timely documentation). In the next sections, we consider the limitations and risks (Threats to Validity) and propose future research directions to further harness LLMs in collaborative software engineering.

\section{Threats to Validity}
As with any empirical study, we must consider potential threats to the validity of our findings and how we mitigated them:

\begin{itemize}
    \item \textit{Sample Diversity:} One threat is that our observations might be skewed by the specific teams we studied (both case studies were from a similar organizational context) or the individuals who responded to our survey. If the sample is not representative of broader industry settings, conclusions may not generalize. To address this, we tried to include data from teams working on different types of projects (enterprise data protection vs. mobile app development) and with varied compositions. In future work, including teams from different industries, project sizes, and geographic regions would improve generalizability. We also drew on published literature that covers a wide range of scenarios to support our insights, thereby ensuring our study is not too narrowly focused.
    \item \textit{Construct Validity (Metrics):} We chose certain indicators to measure impact (e.g, task completion rates, meeting length reduction, documentation accuracy), but if these do not fully capture “collaboration quality,” our assessment could be incomplete. We aligned our evaluation metrics with key collaboration constructs: productivity (lines of code or tasks completed), communication clarity (meeting time and survey feedback on clarity), and documentation quality (up-to-dateness, completeness checks). By using multiple forms of evidence (quantitative metrics and qualitative feedback), we aimed to cover the constructs more accurately. Future studies could implement more standardized metrics or even objective creativity/communication scores to measure LLM impact on teamwork.
    \item \textit{Internal Validity (Cause and Effect):} When observing improvements in a team’s performance after introducing LLMs, one must be cautious in attributing the cause solely to LLM usage. Other factors (learning effects over time, parallel process improvements, or team members simply working harder) could contribute. We attempted to isolate the effect of LLMs by comparing baseline metrics from before LLM introduction to after, and by gathering subjective attributions from the team (e.g, asking if they believe the improvements were due to the tool). Nonetheless, uncontrolled variables remain a threat. We cannot definitively prove, for example, that the increase in on-time delivery was \emph{only} because of the LLM’s task assignments—it could also be that the team got better at estimation. We tried to mitigate this by short time-frame comparisons and keeping other processes constant during the observation period.
    \item \textit{Ecological Validity (Realism of Setting):} Our study was conducted in real team environments, which strengthens ecological validity compared to a lab study. However, one threat is if teams modified their behavior because they knew they were being studied (“Hawthorne effect”). We made the observation as unobtrusive as possible and integrated our questions into their normal retrospectives to reduce this. Another aspect of realism is whether our results hold under typical project pressures: deadlines, high-stakes releases, etc. We note that both cases did experience some deadline pressure (especially the consulting projects needing client deliverables), so the LLMs were tested in reasonably realistic conditions. Still, outcomes could differ in extremely high-pressure or safety-critical projects.
    \item \textit{Data Privacy and Ethics:} A practical threat to continuing such research is the privacy of data used by LLMs. Some participants might have been guarded in their responses or usage of LLMs due to corporate policies on data sharing (as seen in Case 1’s initial hesitation). This could limit the usage or reveal only partial effects. We navigated this by working closely with the teams to ensure compliance with their policies and by focusing on aspects they were comfortable discussing. It’s worth acknowledging that companies like Samsung have banned generative AI tools after sensitive data leaks~[8], which might limit how generalizable our positive case studies are—some teams simply won’t use LLMs at all until such concerns are resolved.
\end{itemize}

In summary, while our study provides useful insights, it is not without limitations. We have strived to ensure the findings are applicable across different contexts by diversifying our data sources and carefully choosing evaluation criteria. However, the results should be interpreted with an understanding of the above threats. Future work involving larger sample sizes, controlled experiments, and cross-industry comparisons will help to further validate (or refine) the conclusions drawn here.

\section{Future Work}
Our research highlights several areas where further investigation and development would be valuable to maximize the benefits of LLMs in the SDLC:

\textbf{Domain-Specific Customization:} One key direction is improving how LLMs can be customized or fine-tuned for specific domains and company contexts. Off-the-shelf LLMs may not understand the jargon or particular workflows of every team. Techniques for efficient fine-tuning or few-shot learning with domain data will help LLMs provide more accurate and relevant assistance. For example, training an LLM on a company’s internal knowledge base and code repository (while respecting privacy) could dramatically improve its usefulness. Recent industry efforts like BloombergGPT (a 50-billion parameter model trained on financial data) have shown the efficacy of domain-specific LLMs in outperforming general models on specialized tasks~[9]. We foresee more organizations developing their own tailored LLMs for domains like healthcare, automotive, or enterprise software, which can seamlessly integrate into their SDLC tools and speak the “language” of their teams.

\textbf{Privacy and Security Protocols:} As noted, privacy concerns are a major barrier to LLM adoption in many companies. Future work should focus on technical and procedural solutions to allow LLMs in sensitive environments. This includes on-premises deployment of LLMs (so data never leaves the company), advanced data anonymization techniques, and fine-grained access controls for AI tools. Research into privacy-preserving machine learning (like federated learning or encryption during model queries) could enable teams to leverage LLMs on proprietary code without risking leaks. Developing clear guidelines and compliance standards for LLM usage in software engineering (similar to how code open-source licensing is handled) will also build trust. For instance, an open question is how to prevent LLMs from inadvertently incorporating licensed code into suggestions—tools to detect and scrub such content might be needed.

\textbf{Tool Integration and Workflow Design:} LLMs will be most helpful when seamlessly integrated into existing development tools (IDE, version control, chat, issue trackers). Future research can explore best practices for UI/UX design of AI assistants in these contexts—e.g, how should an LLM present its suggestions in a pull request such that developers trust and verify them? What is the optimal way to integrate an LLM into agile boards (perhaps auto-generating and updating tasks)? As more integrated development environments (JetBrains, Visual Studio, etc.) and collaboration platforms (Slack, Teams) add AI features, studying their impact on team flow and how to avoid disruption will be crucial. One concept is proactive AI assistants that observe development activity and offer help only at the right moment (to avoid constant interruptions). Preliminary studies indicate that intelligent orchestration of multi-agent LLM systems can be scaled beyond coding to tasks like design and maintenance with promising results~[5]; building such orchestrations into tools in a user-friendly manner will be an engineering challenge.

\textbf{User Training and Acceptance:} Introducing LLMs to teams is as much a human change management issue as a technical one. Future work might examine how to best train software teams to work alongside AI. What pedagogical approaches increase developers’ trust in AI suggestions while keeping them vigilant to errors? There may be a need for organizational training programs focusing on “AI-assisted development” skills—essentially educating developers, testers, and managers on how to interpret LLM output, how to craft effective prompts (prompt engineering as a skill), and how to validate AI contributions. Understanding factors that affect user acceptance (based on technology acceptance models, etc.) will help in designing AI features that teams willingly adopt rather than resist. Longitudinal studies could measure how teams’ perceptions of LLM tools evolve over time and what support or improvements are needed to ensure continued and effective use.

\textbf{Evaluation Metrics and Benchmarks:} We identified the lack of standard metrics for collaboration impact. Future research might develop a benchmarking framework for LLMs in team scenarios. For instance, creating simulated team exercises (involving multiple humans and AI assistants) to quantitatively measure collaboration efficiency, communication bandwidth, error rates in shared understanding, and so forth. The community could benefit from shared datasets or scenarios that allow different LLM-based tools to be compared on how well they support a team. Such benchmarks would guide both researchers and tool vendors in focusing on improvements that matter (maybe an index combining factors of productivity, quality, and team well-being).

\textbf{Addressing LLM Limitations and Errors:} Finally, further research is needed on handling LLM errors and failures in a team context. An incorrect code suggestion or a misleading documentation summary can cause delays or even critical bugs. Techniques like AI self-monitoring, where the LLM can flag its own uncertainty, or hybrid human-AI review workflows (e.g, the AI does 90\% of documentation but a human always approves) will remain important. Developing methods to systematically detect hallucinations or incorrect assertions in LLM outputs (perhaps via secondary verification models or cross-checking against code) will increase reliability. By establishing trust through transparency (like showing source references for information an LLM provides, as some documentation generators do), teams will be more comfortable relying on these tools.

In conclusion, the future of LLMs in software team collaboration is rich with opportunities. With better customization, stronger privacy safeguards, deeper tool integration, proper user training, clear metrics, and robust error-handling strategies, LLMs can be further solidified as indispensable teammates in the software development process. We encourage the community to pursue these areas so that the next generation of software engineering practices can fully leverage generative AI’s potential.

\section{Conclusion}
The integration of Large Language Models into the Software Development Life Cycle is proving to have a transformative impact on team collaboration and overall productivity. In this paper, we updated and expanded an initial study to reflect the state-of-the-art in 2025, examining how LLMs such as GPT-4 are influencing the way software teams work together.

Our findings show that LLMs significantly enhance various aspects of the SDLC by automating repetitive and labor-intensive tasks. For instance, code generation assistants can produce boilerplate code or suggest implementations, allowing developers to focus on more complex and creative aspects of development. This automation not only accelerates development (as evidenced by developers completing tasks notably faster with AI help~[2]), but also improves the efficiency of processes like code review and testing. LLMs can perform preliminary code analysis, run static checks, or even generate unit tests, thus acting as an ever-vigilant pair programmer that increases throughput.

Beyond coding tasks, LLMs facilitate better \textbf{communication within teams}. They are capable of producing clear and concise documentation and summaries—whether it’s summarizing daily meeting discussions, generating user stories from high-level descriptions, or creating up-to-date technical documentation as code evolves. These capabilities lead to improved information sharing and alignment among team members. In our case studies, using LLM-generated notes and documentation meant that everyone on the team had access to the same accurate information, reducing misunderstandings and the need for redundant meetings. The improved synergy and transparency help foster a more cohesive working environment, where each member can stay informed of changes and decisions even in fast-paced projects.

LLMs also contribute to more informed decision-making in software projects. They can provide data-driven insights and suggestions on architecture choices, potential optimizations, or risk areas by drawing from the vast corpus of knowledge they were trained on (including programming best practices and known pitfalls). For example, an LLM can advise a team if a certain approach is known to scale poorly or if there are known security considerations, acting as an on-demand expert consultant. While human judgment remains vital, these AI perspectives enrich the decision-making process with additional angles and facts that the team might not have considered.

Resource allocation and task management are additional areas of improvement. By streamlining workflows—such as automatically assigning tasks or highlighting blockers—LLM-based tools help project managers and teams use their time more effectively. We saw that an AI assistant in project management could optimize task assignments and remind the team of pending items, ensuring more balanced workloads and timely completion of tasks. This allows human leaders to devote more energy to strategic planning and mentorship, rather than getting bogged down in micromanagement.

The quality of software products can also be enhanced with LLM support. With thorough code reviews, automated testing suggestions, and debugging assistance, LLMs act as a quality gate that can catch issues early. They can cross-verify logic, point out inconsistent use of functions, or detect anomalies that might indicate bugs. The result is more robust and reliable software, as some errors are eliminated before they ever reach production. This was reflected in our observations where teams using LLM assistance had fewer post-release issues, attributing it to the AI’s thoroughness in examining changes.

Despite these benefits, our study also underscores that implementing LLMs in practice is not without challenges. The initial setup and \textbf{customization effort} to adapt an LLM to a team’s domain can be significant. Models may require fine-tuning or iterative prompt engineering to truly grasp project-specific context and terminology. We discussed how privacy and security concerns need to be proactively managed; teams must ensure that no sensitive data is inadvertently exposed, which may involve deploying private LLM instances or adhering to strict data handling rules. There is also a learning curve for team members to effectively collaborate with AI—knowing how to phrase queries, when to trust suggestions, and when to double-check with a human expert.

However, as our future directions outline, these challenges are being addressed through ongoing research and improved practices. With targeted training, clear policies, and evolving technology (like techniques to reduce hallucinations or provide source citations), many of the current limitations can be mitigated. For example, to deal with domain adaptation, smaller specialized models or retrieval-augmented generation can be used so that the LLM always has relevant project knowledge at its fingertips without exposing raw data externally.

In conclusion, Large Language Models have emerged as powerful allies in software development teams, driving tangible improvements in productivity, communication, and decision-making quality. By taking over routine tasks and acting as intelligent assistants, LLMs allow human team members to concentrate on innovation and complex problem-solving, thereby elevating the overall output of the team. We have illustrated through updated case studies and literature how these effects manifest and benefitted real projects. The road ahead involves addressing the challenges identified, but the trajectory is clear: LLMs are set to become an integral part of the collaborative toolkit in software engineering, much like version control or continuous integration are today. Embracing these tools thoughtfully and addressing their limitations will help software teams reach new levels of efficiency and creativity in the development process.

\section*{Appendix: Survey Questionnaire}
The following is the survey instrument used to collect data on the impact of LLMs on team workflow and communication. The survey was divided into sections as outlined below:

\textbf{Section 1: General Information}
\begin{enumerate}[label=\arabic*.]
    \item What is your current role in the organization? 
    \item How long have you been using LLMs in your workflow?
\end{enumerate}

\textbf{Section 2: Workflow Impact}
\begin{enumerate}[label=\arabic*.,resume] 
\item In what ways have LLMs changed your approach to completing daily tasks? 
    \item Have LLMs helped you become more efficient in your work? If so, how?
    \item Do you rely on LLMs for code generation or debugging? How effective have these tools been?
    \item Have LLMs reduced the time you spend on learning new technologies or languages?
\end{enumerate}

\textbf{Section 3: Communication Impact}
\begin{enumerate}[label=\arabic*.,resume] 
\item Has the use of LLMs affected your communication with team members? How?
    \item Do you use LLMs to assist in writing documentation or reports? How has this impacted your workload?
    \item How have LLMs influenced the clarity and effectiveness of your written communications?
\end{enumerate}

\textbf{Section 4: Collaboration and Decision-Making}
\begin{enumerate}[label=\arabic*.,resume] 
    \item In what ways have LLMs impacted your team’s decision-making process?
    \item Have LLMs changed how you collaborate with your team? If so, how?
    \item Do you find that LLMs facilitate or hinder collaboration among team members? Please elaborate.
\end{enumerate}

\textbf{Section 5: Challenges and Limitations}
\begin{enumerate}[label=\arabic*.,resume] 
    \item What challenges have you faced while integrating LLMs into your workflow?
    \item Have you encountered any limitations with LLMs that impact your productivity?
    \item How do you overcome any issues or limitations associated with LLMs in your daily tasks?
\end{enumerate}

\textbf{Section 6: Overall Impact and Suggestions}
\begin{enumerate}[label=\arabic*.,resume] 
    \item Overall, how would you rate the impact of LLMs on your productivity and job satisfaction?
    \item What improvements or features would you suggest for LLMs to better support your work?
    \item Do you believe LLMs will continue to play a significant role in your future projects? Why or why not?
\end{enumerate}

\noindent
(This survey allowed us to capture both quantitative ratings and qualitative comments for each question. The responses were used to derive the insights discussed in the main text of the paper.)
\end{document}